\documentclass[twocolumn,superscriptaddress,aps,showpacs,amsmath,footinbib,bibnotes]{revtex4-1}
\pdfoutput=1
\usepackage{graphicx}
\usepackage{amssymb,amsmath}
\usepackage{textcomp} 
\usepackage[bold]{hhtensor}
\usepackage{natbib}
\usepackage{lipsum}
\usepackage{bm}
\usepackage{hyperref}
\usepackage{multirow}
\usepackage{float}
\usepackage{amsmath}
\usepackage{amssymb}
\usepackage{stmaryrd}
\usepackage{mathrsfs}
\usepackage[dvipsnames]{xcolor}
\usepackage{makecell}
\usepackage[flushleft]{threeparttable}

\hypersetup{
	colorlinks,
	linkcolor={red!80!black},
	citecolor={Blue},
	urlcolor={MidnightBlue}
}

\renewcommand \bf[1]{\textbf{#1}}

\begin{document}

\author{Prashanta Kharel}
\affiliation{HyperLight, 501 Massachusetts Avenue, Cambridge, Massachusetts 02139, USA}
\author{Christian Reimer}
\affiliation{HyperLight, 501 Massachusetts Avenue, Cambridge, Massachusetts 02139, USA}
\author{Kevin Luke}
\affiliation{HyperLight, 501 Massachusetts Avenue, Cambridge, Massachusetts 02139, USA}
\author{Lingyan He}
\affiliation{HyperLight, 501 Massachusetts Avenue, Cambridge, Massachusetts 02139, USA}
\author{Mian Zhang}\email{mian@hyperlightcorp.com}
\affiliation{HyperLight, 501 Massachusetts Avenue, Cambridge, Massachusetts 02139, USA}

\date{\today}
\title{Breaking voltage-bandwidth limits in integrated lithium niobate modulators using micro-structured electrodes}

\begin{abstract}
Electro-optic modulators with low voltage and large bandwidth are crucial for both analog and digital communications. Recently, thin-film lithium niobate modulators have enable dramatic performance improvements by reducing the required modulation voltage while maintaining high bandwidths. However, the reduced electrode gaps in such modulators leads to significantly higher microwave losses, which limit electro-optic performance at high frequencies. Here we overcome this limitation and achieve a record combination of low RF half-wave voltage of 1.3 V while maintaining electro-optic response with 1.8-dB roll-off at 50 GHz. This demonstration represents a significant improvement in voltage-bandwidth limit, one that is comparable to that achieved when switching from legacy bulk to thin-film lithium niobate modulators. Leveraging the low-loss electrode geometry, we show that sub-volt modulators with $>$ 100 GHz bandwidth can be enabled.
\end{abstract}
\maketitle

\section{Introduction}

Low voltage, broadband and high signal quality electro-optic (EO) modulators are paramount to applications spanning from radiofrequency (RF) analog links to digital optical communication networks. Today, most EO modulators require high voltage drives at microwave frequencies $>$ 50 GHz because of reduced high frequency performances. This high RF voltage requires high-speed and high-gain electrical amplifiers which pose challenges for power consumption, linearity, signal-to-noise ratio and cost. As the demands for high baud rate for digital communication and higher carrier frequency for analog link continue to grow, aforementioned challenges only exacerbate at even higher microwave frequencies (e.g. $>$ 100 GHz), as modulators’ efficiency continues to decrease, electronic amplifiers gain also diminishes at higher microwave frequencies.

The challenge of achieving low voltage at high RF frequencies is universal across different photonics platforms, considering the stringent requirement of simultaneously retaining other desired properties including good linearity, low insertion loss, high extinction ratio and high-power handling ability. For traditional lithium niobate (LN) modulators based on ion-indiffusion and proton exchange, a half-wave voltage $V_\pi \sim 3.5$ V, which is defined as the voltage needed to switch the modulation from maximum to the nearest minimum transmission, is typically needed at low RF frequencies (e.g. 1 GHz). The EO response is attenuated by 3-6 dB for $>$ 50 GHz \cite{wooten_review_2000}. This translates into a RF voltage requirement as high as $>$ 7 V for modulation frequencies $>$ 50 GHz. On integrated platforms such as silicon, modulators typically have $V_\pi \sim$ 4-6 V and 35 GHz bandwidth \cite{zhou_silicon_2019,li_silicon_2018}. Extending this performance to 50 GHz and beyond also points to very high voltage ($>$ 10 V) requirement. Indium phosphide (InP) modulators have achieved better voltage and bandwidth performances. For example, $V_\pi$ = 1.5 V and 80 GHz 3-dB bandwidth has been achieved on a differential RF drive architecture \cite{ogiso_80-ghz_2020}. In addition, sub-volt single-drive modulators have also been achieved with 67 GHz 6-dB bandwidth on III-V platforms \cite{dogru_077-v_2014}, but the extinction ratio was limited to 3 dB and drive voltage needed to be increased in order to accommodate higher optical power \cite{bhasker_low_2020}. Organic polymer modulators have shown excellent voltage-bandwidth performances \cite{kieninger_ultra-high_2018} but often the performances need to be compromised in order to improve stability for practical uses \cite{kieninger_demonstration_2018}. Plasmonic-organic hybrid can provide extremely high bandwidths, although modulation voltage and on-chip insertion loss remain relatively high \cite{burla_500_2019}. 
 \begin{figure*}
    \centering
    \includegraphics[width=0.9\textwidth]{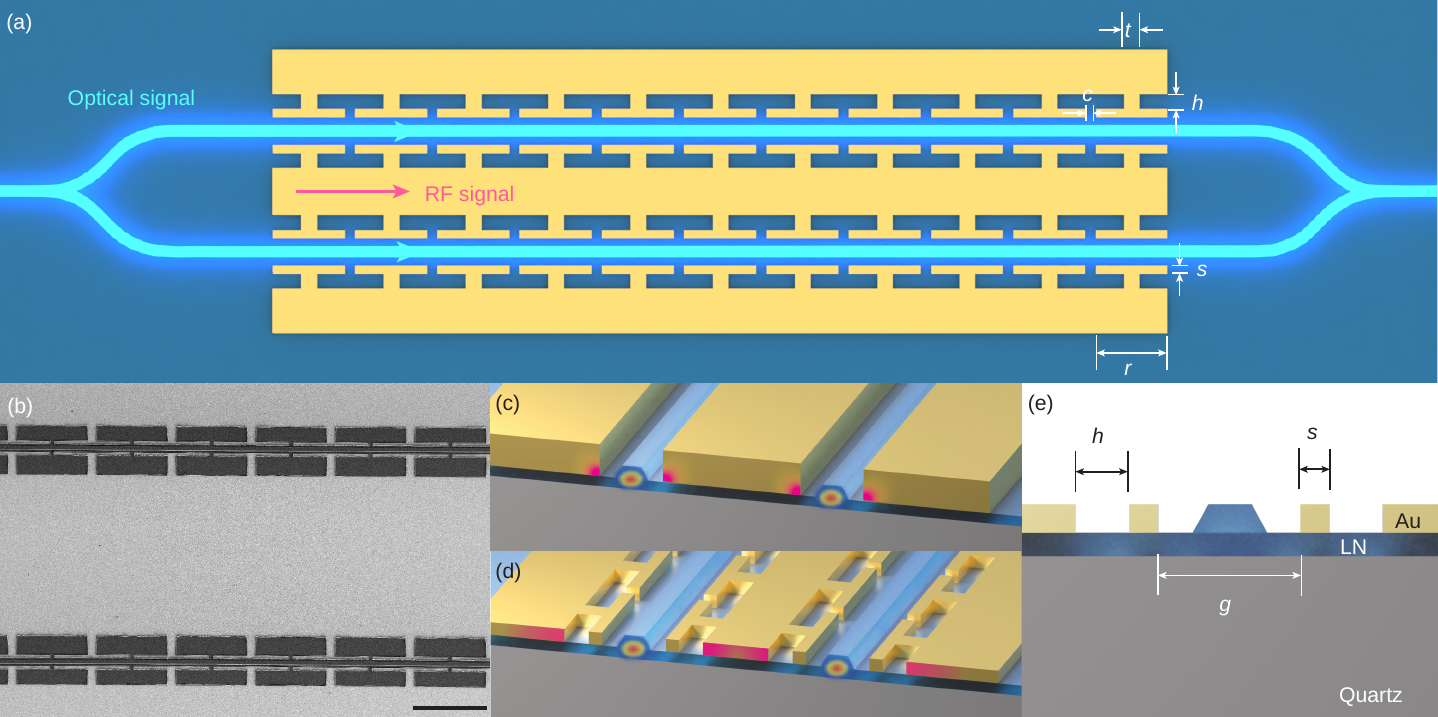}
    \caption{Low-voltage high bandwidth traveling-wave integrated lithium niobate (LN) modulator with segmented electrodes. (a) Artistic top view of the modulator design (not to scale) where RF signal co-propagates with the optical signal . (b) Scanning electron microscope image of the fabricated device. Scale bar: 50 $\mu$m. (c) Artistic angled view of a regular electrode design. Current (red) crowds at edges of the conductors. (d) Artistic angled view of a segmented design. Current crowds less and distribute more uniformly. (e) Cross sectional view of the phase shifter. Design parameters $(g,h,s,t,r,c)= (5,6,2,6,45,5) \mu$m.}
    \label{fig1}
 \end{figure*}
\section{Segmented traveling wave LN modulator design}
Thin-film LN modulators emerged recently as a strong contender for next generation low-voltage and high bandwidth EO modulators. This is because thin-film LN modulators offer significantly improved voltage-bandwidth performance over legacy LN platforms all while preserving key LN material advantages such as linear response, high extinction ratio, high optical power handling ability and low on-chip insertion loss. However, for thin-film LN modulators, $V_\pi$, especially at frequencies $>$ 50 GHz, remains $>$ 3 V \cite{wang_integrated_2018,xu_high-performance_2020,ahmed_high-efficiency_2020}. Extrapolating to 100 GHz shows an expected $V_\pi >$ 4 V. Such performances have also been corroborated theoretically as being close to the traditional design limit \cite{honardoost_towards_2019,honardoost_high-speed_2018,safian_foundry-compatible_2020,peter_orlando_design_2020}. While the voltage-bandwidth performances are a significant improvement over legacy bulk LN technologies, the level of RF voltages achieved in existing thin-film LN modulators at $>$50 GHz are still very high for typical electronics drivers. For example, a CMOS driver with a 100 GHz analog bandwidth, would produce $\sim$ 0.5 V output voltage at high frequencies \cite{chen_all-electronic_2017}. Therefore, a lower voltage modulator would dramatically improve device performance such as energy consumption, sensitivity, and noise figures which all scales quadratically with RF $V_\pi$ \cite{urick_fundamentals_2015}.

Here we break the voltage-bandwidth trade-off limit in integrated LN modulators using micro-structured electrodes that dramatically reduce microwave losses while preserving EO modulation efficiency as well as other desired properties including high extinction ratio and low on-chip optical loss. We experimentally demonstrate a single drive EO modulator with $V_{\pi,\textrm{1GHz}} = 1.3$ V and $V_{\pi,\textrm{50GHz}} = 1.6$ (EO roll-off of 1.8 dB). We further show that this design can be adapted to achieve sub-volt modulators with $>$ 100 GHz 3-dB EO bandwidth. At the same time, we maintain a on-chip loss of $<$ 1 dB and high extinction ratio of 20 dB. Notably, the performance gain achieved using the micro-structured electrode design over regular electrodes on thin-film LN is comparable to the improvement attained when transitioning from bulk to thin-film LN modulators.

\begin{table*}[htbp]
\begin{threeparttable}
\centering
\caption{\bf Comparison of simulated performance for various electrode designs}
\begin{tabular}{ccccccc}
\hline
\makecell{Electrode design \\ (substrate)} & \makecell{Signal width \\ ($\mu$m)} & \makecell{Gap \\ ($\mu$m)} & \makecell{$V_\pi\cdot L$\\ (V$\cdot$cm)} & $Z$ ($\Omega$) & \makecell{RF loss \\ (dB cm$^{-1}$GHz$^{-1/2}$)} & \makecell{Est.3-dB EO \\ bandwidth for \\ $V_{\pi,\textrm{DC}}=1$V (GHz)}\\
\hline
Regular (Si) & 30 & 5 & 2.1 & 39 & 0.75 & 27 \\
Regular (Si) & 30 & 7 & 2.6 & 45 & 0.64 & 25 \\
Regular (Si) & 30 & 10 & 3.7 & 48 & 0.57 & 17 \\
Regular (Q) & 100 & 5 & 2.1 & 41 & 0.37 & 13*\\
Segmented (Q) & 100 & 5 & 2.1 & 42 & 0.21 & 228 \\

\hline
\end{tabular}
    \begin{tablenotes}
      \small
      \item ~Q: quartz. $Z$: impedance. 
      \item$^*$Velocity mismatch assumed.
    \end{tablenotes}

  \label{table1}
  \end{threeparttable}
\end{table*}
The EO modulator employs traveling wave design on a x-cut thin-film LN-on-insulator (LNOI) platform, where an input light is split into two arms of a Mach-Zehnder interferometer (MZI) and co-propagate with a microwave drive signal in a transmission line electrode \cite{wang_integrated_2018}. The traveling microwave signal modulates the light throughout the lengths of the electrodes in a push-pull configuration, inducing a phase advance in one arm and phase delay in the other. In principle, traveling wave modulators in LN with a long enough electrode can achieve THz bandwidth and millivolt driving voltage since Pockels effect takes place on femtosecond time scale \cite{boyd_nonlinear_2008}. In practice, bandwidth and voltage performance is limited by three key factors: 1) microwave loss in the transmission line causing driving voltage to be attenuated over length of the electrode; 2) finite velocity mismatch between the electrical and optical traveling signal causing modulation to cease accumulating; and 3) design trade-offs that reduce modulation efficiency per unit length which could require even longer devices to achieve a low voltage thus further limiting the bandwidth. 

In other words, to achieve high-bandwidth and low-voltage operations on LN, the speed of the electrical signal should match that of the optical signal, microwave loss (i.e. the attenuation of electrical current flowing in the direction of the traveling wave) should be low, and the electric field should be strong between the electrode gaps so that the modulation is phase matched and can efficiently accumulate along the traveling wave direction.

While modulation efficiency can be increased and velocity matching is more readily maintained in thin-film LN modulators when compare to legacy bulk LN modulators \cite{rao_compact_2018}, microwave loss unfortunately increases significantly in integrated LN electrodes in comparison to bulk LN designs \cite{honardoost_towards_2019,rao_compact_2018,doi_advanced_2006}. Microwave electrode losses originate from two sources: substrate absorption loss (e.g. LN, Si) and ohmic conductor loss from finite resistivity of metals, with the latter being the dominant loss mechanism in existing thin-film LN modulators. This is because the smaller electrode gaps in thin-film LN modulators, enabled by high confinement optical waveguides, improve key efficiency metrics like half-wave voltage length product ($V_\pi\cdot L$) at the expense of dramatically increased ohmic loss. The narrow metal gap, on the order of a few micrometers, causes the electrical current to crowd close to the gap from the largely increased capacitance, effectively reducing conductor area and thus increasing RF loss. As a result, non-ideal trade-offs such as increasing the electrode gap which lowers modulation efficiency (increases $V_\pi\cdot L$) and/or reducing electrode length have been predicted in order to maintain flat RF responses (Table 1) \cite{honardoost_towards_2019,honardoost_high-speed_2018,peter_orlando_design_2020}. Such design trade-offs lead to underutilization of a tightly confined optical mode for efficient EO modulation on on integrated LN platform. 

In contrast, our design maintains the high modulation efficiency (low $V_\pi\cdot L$) on thin-film LN platform while reducing the electrode losses and maintaining velocity matching condition. To achieve this, we employ a traveling wave electrode with micro-structures to control the flow of currents. The micro-structured electrodes consist of rectangular channel electrode regions like conventional coplanar transmission line designs and segments extending out from the main electrode (Fig. \ref{fig1}a,b). The segments prevent electrical current from flowing in the closest gap region, while allowing the current to be distributed more uniformly in the wide channel region. As a result, the effective conductor size is increased and the ohmic loss in the electrode is reduced without having to increase the gap ($g$) between the electrodes (Fig. \ref{fig1}c,d). 

Such micro-structured electrodes (also known as segmented electrodes) have been used previously on semiconductor substrates to help velocity matching \cite{jaehyuk_novel_2005} and also moderately improve conductor loss in III-V and silicon modulators \cite{shin_conductor_2011,ding_high-speed_2014}. On insulators with high permittivity such as LN ($\epsilon_\textrm{LN} \sim$ 30), such structures have not been previously implemented to improve RF performances. This is because while the extent of improvement of microwave loss with such electrode design on insulator was unknown, a collateral and deleterious effect of the segmented design is that the microwave velocity would be significantly reduced relative to a rectangular electrode design due to the increased capacitance per unit length of the transmission line from the segments. For traditional LN substrate, this slow wave effect would be detrimental for velocity matching since RF velocity is already slower than optical group velocity to begin with \cite{wooten_review_2000}. On thin-film LN on silicon substrate, the slow wave effect is also undesirable since silicon substrate with a reasonably thick silicon dioxide insulating layer can already provide a good velocity match between light and microwave \cite{wang_integrated_2018,rao_compact_2018}.

Here we instead take advantage of the slow wave effects by using a quartz substrate, which has a near ideal microwave properties with low permittivity ($\epsilon_\textrm{Qz} \sim$ 4.5) and microwave absorption tangent $< 10^{-4}$ \cite{geyer_microwave_1995}. On a LN thin film with quartz substrate, one would obtain a RF index $\sim$ 1.8 for conventional rectangular electrodes, which is significantly lower than the typical optical group index of $\sim$ 2.2. While this velocity mismatch would be detrimental for high speed operations for regular electrode designs \cite{stenger_low_2017,mercante_thin_2018}, we utilizes it to our advantage to design segments that pushes the current away from narrow gaps. We show that our segmented design can be used to achieve RF index of $\sim$ 2.2, which can be fine-tuned by the dimensions of the segments to match the optical group velocity in LN precisely. Importantly, as opposed to marginal RF loss improvements in other semiconductor platforms, the segmented electrode design in thin-film LN reduces the RF loss substantially.

 \begin{figure*}
  \includegraphics[width=0.9\textwidth]{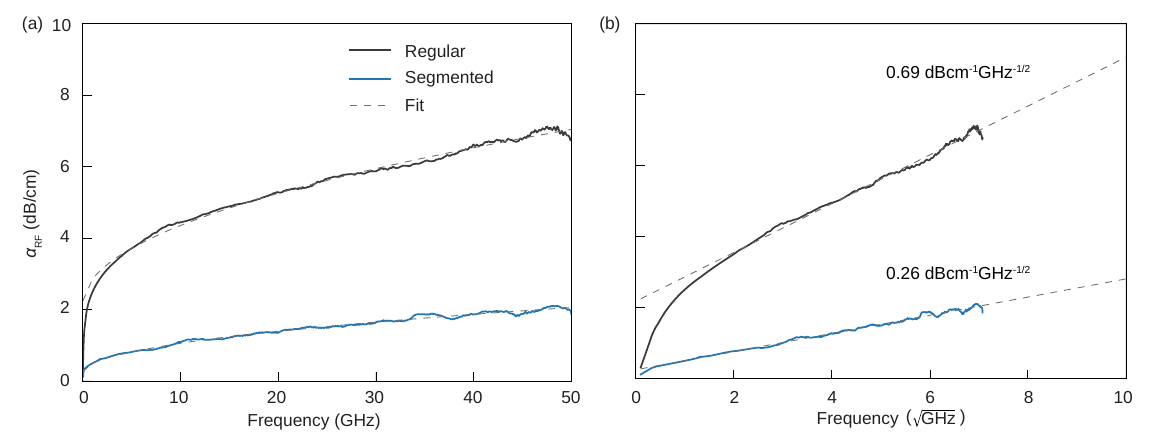}
    \caption{Measured RF loss of a 10-mm long regular electrode on LN-on-silicon and segmented electrode on LN-on-quartz with square-root fitting on (a) linear frequency axis (b) square-root frequency axis.
    }
    \label{fig2}
 \end{figure*}
 
We simulated the performance of segmented electrodes and regular electrodes using finite element methods (HFSS) and compare the results in Table \ref{table1}. For regular electrodes on silicon, the film stacks used were LN: 600 nm, SiO$_2$: 2 $\mu$m and Si: 500 $\mu$m. For electrodes on a quartz substrate the film stacks wereare LN: 600 nm  and quartz: 500 $\mu$m. The segmented electrode uses a parameter set of $(h,s,t,r,c) = (6,2,6,45,5)~\mu$m as defined in Fig. \ref{fig1}. It is clear from Table 1. that using segmented electrodes, the modulator 3-dB bandwidth can be improved by more than 8 times to 228 GHz in comparison to existing approaches for the same designed DC $V_\pi$.

\section{Measurements and analysis}

We fabricated LN modulator device using a wafer (NanoLN) consisted of a 600 nm thick x-cut LN thin-film on a 500 $\mu$m thick quartz handle. We patterned the optical device with electron beam lithography and etched 350 nm of LN film to define integrated waveguides using a previously reported method \cite{zhang_monolithic_2017}. We then patterned 800 nm thick gold electrodes with a lift-off process. The device was also cladded with 1 $\mu$m thick silicon dioxide, deposited by chemical vapor deposition.

We show that the segmented electrode completely transformed the RF performance while preserving other desired modulator performances. We measured the electrical loss on the segmented electrode using a 50-$\Omega$ vector network analyzer (VNA) with all the RF components up to the device de-embedded. We obtain RF loss of 2 dB/cm in the segmented electrode at 50 GHz in comparison to 7 dB/cm in regular electrode design with identical gold electrode thickness of 800 nm and electrode gap $g = 5 \mu$m (Fig. \ref{fig2}a). We also measured a RF phase index of 2.23 which agrees with our simulation. 

 \begin{figure*}
  \includegraphics[width=0.9\textwidth]{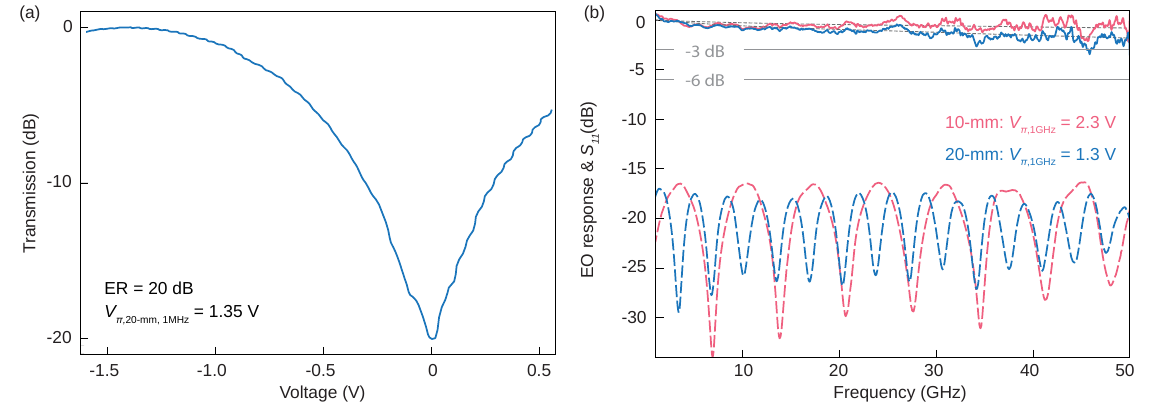}
    \caption{Electro-optic performance of segmented LN modulators (a) Measured DC $V_\pi$ and extinction for a 20-mm long modulator. (b) Measured EO response and electrical reflection ($S_{11}$) referenced to RF $V_\pi$ at 1 GHz for a 10-mm and a 20-mm long modulator. The response is normalized to photodiode electric signal, i.e. $V_{\pi,\textrm{3-dB}} \sim 1.4 V_{\pi,\textrm{1GHz}}$ and $V_{\pi,\textrm{6-dB}} \sim 2 V_{\pi,\textrm{1GHz}}$}
    \label{fig3}
 \end{figure*}
 
Our measured results have excellent agreement with theory. Ohmic loss in the electrode $\alpha_\textrm{RF}$, is $\propto Lf^{1/2}$ as a result of the skin effect in metal \cite{haxha_bandwidth_2003}, where $L$ is the length of the electrode and f is microwave frequency. We measure Regular electrodes on thin-film LN have $\alpha_\textrm{RF,reg}$ = 0.69 dB cm$^{-1}$GHz$^{-1/2}$ compare to $\alpha_\textrm{RF,seg}$ = 0.26 dB cm$^{-1}$GHz$^{-1/2}$ for the segmented electrodes. The square root dependence of the loss is clear from Fig. \ref{fig2}b, corroborating the assumption that ohmic loss is the dominate source of RF attenuation. Note that microwave absorption loss has a linear dependence with microwave frequency so it would show up as a superlinear contribution in Fig. \ref{fig2}b at high frequencies. Our results show that the linear loss is still too small in comparison to conductor loss up to 50 GHz. Based on our fitting, we estimate an upper limit for RF absorption on LN on quartz substrate of 0.007 dB cm$^{-1}$GHz$^{-1}$, which translates to less than 0.35 dB/cm at 50 GHz and 0.7 dB/cm at 100 GHz. Using a loss tangent of 0.004 for LN bulk crystals \cite{lee_dielectric_2001}, our simulation produced absorption loss of 0.003 dB cm$^{-1}$GHz$^{-1}$, which agrees well with the limited estimated by our measurements.

 \begin{figure*}
  \includegraphics[width=0.9\textwidth]{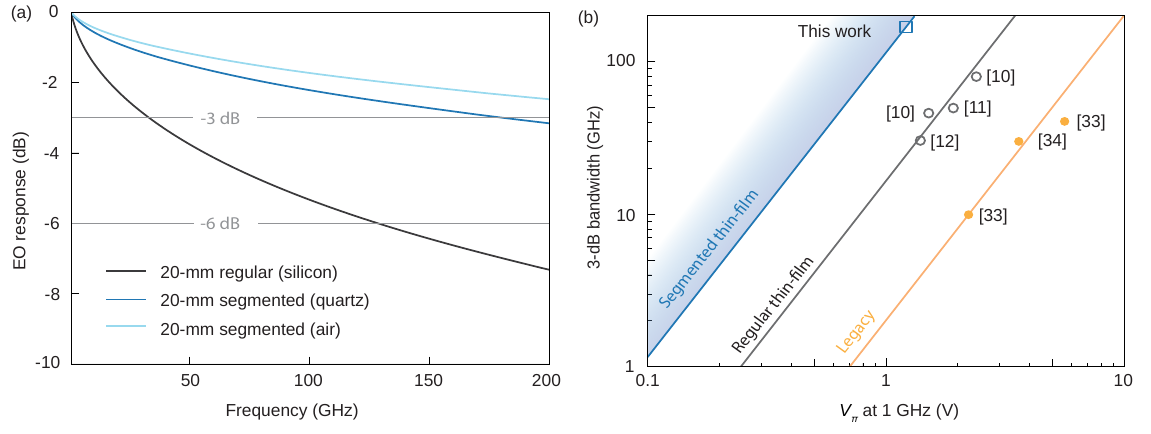}
    \caption{Comparison of micro-structured modulator performance to prior designs. (a) Predicted performances of 20-mm long electrode for regular and segmented design up to 200 GHz bandwidth. Material legends indicate the substrate handle. (b) Voltage-bandwidth performance comparison of legacy LN (0.3 dB cm$^{-1}$GHz$^{-1/2}$, 15 V$\cdot$cm), thin-film regular LN (0.69 cm$^{-1}$GHz$^{-1/2}$, 2.1V$\cdot$cm) and thin-film segmented LN modulators (0.26 cm$^{-1}$GHz$^{-1/2}$, 2.3 V$\cdot$cm). The shaded areas correspond to improved design space with other substrates. The bandwidth for this work is extrapolated as shown in (a). }
    \label{fig4}
 \end{figure*}
 
We measured the EO performance of the modulators using a telecom wavelength tunable laser at 1560 nm. We coupled light through a pair of grating couplers with total insertion loss 13 dB and on-chip loss $<$ 1 dB estimated by comparing transmission loss of a pair of gratings couplers with and without the modulator. Majority of the optical loss results from the 6 dB/facet loss of grating couplers due to the absence of a high index substrate. The total insertion loss can be dramatically reduced by using edge couplers \cite{he_low-loss_2019}, or buried metal back reflectors \cite{chen_grating_2017} to $\sim$ 4 dB. We obtained DC $V_\pi$ of 1.35 V on an oscilloscope for a 20-mm long modulator. The extinction of the modulator was measured to be about 20 dB (Fig \ref{fig3}a). 

The ultralow RF loss enabled measured EO responses of only 1.8 dB attenuation for the 20-mm long modulator at 50 GHz comparing to reference $V_\pi$ at 1 GHz (Fig. \ref{fig3}b). We choose to reference the EO roll-off to $V_{\pi,\textrm{1GHz}}$ which is also conventional, since RF $V_\pi$ is overall a better metric for gauging modulator performances. This is because LN modulators at close to DC frequencies are prone to additional slow effects such as photorefractive effect that can lead to over- or underestimation of $V_\pi$. Our measurements at 1 GHz using a Bessel function transform method \cite{nagarajan_technique_1999} indicate $V_{\pi,\textrm{1GHz,20-mm}} = 1.3$ V.  In other words, the RF $V_\pi$ at 50 GHz is a record low of only 1.6 V. We also measured a 10-mm modulator with segmented electrode, which shows only 0.8 dB roll-off at 50 GHz relative to a measured $V_{\pi,\textrm{1GHz,10-mm}} = 2.3$ V. The electrical reflection ($S_{11}$) from the electrode for both cases is maintained below -15 dB. The measured RF $V_\pi$ for 20-mm electrode is higher than expected due to finite resistance of the thin metal electrode, which can be improved with thicker metals.

\section{Discussion}

The flexibility of substrate engineering and micro-structured electrode design on thin-film LN enabled dramatic performance improvement over standard electrode designs on LN-on-silicon substrates. The current segmented design is estimated to work and behave like a lumped element until $\sim$ 300 GHz, where frequency dependent phase-shift is expected to cause velocity mismatch \cite{shin_conductor_2011}. This frequency limit can be readily overcome by using smaller segments structure allowing operation in the THz regimes.

We analyze the bandwidth and voltage trade-offs in the segmented design (this work) over integrated LN modulators. We extrapolate the expected voltage-bandwidth performance based on our measurements up to 200 GHz (Fig \ref{fig4}a). We see that for the same electrode length of 20 mm, the significant reduction of RF loss in segmented design could lead to 180 GHz conductor loss limited 3-dB bandwidth in comparison to regular thin-film modulators that have 40 GHz bandwidth (Fig \ref{fig4}a). The performance of the segmented electrode design could be increased further using an even lower permittivity substrate such as fused silica ($\epsilon_{\textrm{fs}} = 3.8$) or air ($\epsilon_\textrm{air} = 1$). Decrease in substrate permittivity allows longer segments, which further reduces current crowding while still maintaining perfect velocity matching conditions. We estimate that microwave loss 0.15 dB cm$^{-1}$GHz$^{-1/2}$ is within reach for such designs. This improvement in microwave loss should permit modulators with $V_\pi$ of 780 mV at 100 GHz using an electrode length of 25 mm and optimized gap ($V_\pi$ of 650 mV at 1 GHz and $>$ 200 GHz 3-dB bandwidth with respect to 1 GHz). Note that for such low drive voltages designs, while the electrode length is substantial, the tight optical mode confinement in thin-film LN and small lateral size of the electrode permits bending of the electrodes to fit in a small footprint $<$ 10 mm $\times$ 0.5 mm. 

The micro-structured electrodes push integrated LN modulator performance into a completely new performance space. Here we compare the voltage-bandwidth limits of legacy modulators \cite{thorlabs_lithium_2020,eospace_40_2020}, regular thin-film LN modulators \cite{wang_integrated_2018,xu_high-performance_2020} and here-presented segmented thin-film LN modulators (Fig \ref{fig4}b). The voltage-bandwidth performance due to the current crowding effect, typically have 5-7 dB/cm RF loss at 50 GHz (0.7-1 dB cm$^{-1}$GHz$^{-1/2}$), which is much higher than typical RF loss in legacy modulators of $\sim$ 2 dB/cm at 50 GHz ($\sim$ 0.3 dB cm$^{-1}$GHz$^{-1/2}$). Still, thin-film LN modulators out performances legacy designs due to the nearly 5 times reduction in $V_\pi\cdot L$. On segmented electrode platform, we can maintain the $V_\pi\cdot L$ close to the regular electrode and at the same time improve RF loss to $\sim$ 0.2 dB cm$^{-1}$GHz$^{-1/2}$. Lower index substrates allow the loss to be improved further. From Fig. \ref{fig4}b, we can see that the new segmented design leads to a similar performance gain to what regular thin-film design has achieved over legacy bulk LN.

\section{conclusion}
We have demonstrated an integrated LN EO modulator with ultra-flat frequency response and low RF $V_\pi$ using segmented traveling-wave electrode on low permittivity substrate. We show that sub-volt level of microwave driving voltage can be achieved even at frequency $>$ 100 GHz. We believe the significantly improved EO modulation performance in micro-structured thin-film LN modulators will lead to a paradigm shift for both analog and digital ultra-highspeed RF links. For example, for digital applications with sub-volt modulators, high speed electronic drivers may have largely reduced gain-bandwidth requirement or possibly be completely by-passed with the modulators directly driven from electronic processors \cite{li_electronicphotonic_2020}.

\bibliography{referencelongv}
\end{document}